\documentclass[aps,prl,twocolumn,preprintnumbers,
              showpacs,nofootinbib]{revtex4}
\newcommand{\PRE}[1]{}       

\usepackage{bm}
\usepackage{epsfig}

\newcommand{\sigmath}{\sigma_0^{\text{th}}}
\newcommand{\sigmaan}{\sigma_{\text{an}}}

\newcommand{\mev}{\text{MeV}}
\newcommand{\gev}{\text{GeV}}
\newcommand{\tev}{\text{TeV}}

\newcommand{\cm}{\text{cm}}

\newcommand{\km}{\text{km}}
\newcommand{\fm}{\text{fm}}

\newcommand{\s}{\text{s}}

\newcommand{\kpc}{\text{kpc}}

\newcommand{\etal}{{\em et al.}}
\newcommand{\eg}{{\em e.g.}}
\newcommand{\ie}{{\em i.e.}}

\newcommand{\eqref}[1]{Eq.~(\ref{#1})}
\newcommand{\eqsref}[2]{Eqs.~(\ref{#1}) and (\ref{#2})}
\newcommand{\Eqref}[1]{Equation~(\ref{#1})}

\newcommand{\figref}[1]{Fig.~\ref{fig:#1}}

\newcommand{\ssection}[1]{{\em #1}.}
\newcommand{\vrel}{v_{\text{rel}}}

\begin{document}

\preprint{UCI-TR-2009-12}

\title{ \PRE{\vspace*{1.5in}} Halo Shape and Relic Density Exclusions
of Sommerfeld-Enhanced Dark Matter Explanations of Cosmic Ray Excesses
\PRE{\vspace*{0.3in}} }

\author{Jonathan L.~Feng, Manoj Kaplinghat, and Hai-Bo Yu}
\affiliation{Department of Physics and Astronomy, University of
California, Irvine, California 92697, USA \PRE{\vspace*{.5in}} }

\date{November 2009}

\begin{abstract}
\PRE{\vspace*{.3in}} Dark matter with Sommerfeld-enhanced annihilation
has been proposed to explain observed cosmic ray positron excesses in
the 10 GeV to TeV energy range. We show that the required enhancement
implies thermal relic densities that are too small to be all of dark
matter. We also show that the dark matter is sufficiently
self-interacting that observations of elliptical galactic dark matter
halos exclude large Sommerfeld enhancement for light force carriers.
Resonant Sommerfeld enhancement does not modify these conclusions, and
the astrophysical boosts required to resolve these discrepancies are
disfavored, especially when significant self-interactions suppress
halo substructure.
\end{abstract}


\maketitle

\ssection{Introduction}
Recently PAMELA~\cite{Adriani:2008zr}, ATIC~\cite{:2008zzr},
Fermi~\cite{Abdo:2009zk}, and HESS~\cite{Aharonian:2009ah} have
observed the spectrum of cosmic ray positrons with energies between 10
GeV and a few TeV.  Some of the data show excesses over background
expectations~\cite{Strong:2009xj}.  The excesses have plausible
astrophysical explanations~\cite{Hooper:2008kg,Dado:2009ux}.  At
the same time, signals from many dark matter candidates are expected
in this energy range, and this possibility has not escaped attention.

By far the most researched possibility is that the observed positrons
are produced by dark matter annihilation.  If dark matter $X$ is a
thermal relic, the relic density implies that its thermally-averaged
annihilation cross section times relative velocity at freeze out is
$\langle \sigmaan \vrel \rangle \approx \sigmath \equiv 3\times
10^{-26}~\cm^3/\s$.  Unfortunately, if this is the annihilation cross
section now, the resulting signal is too small by two to three orders
of magnitude to explain the observed cosmic ray excesses.

A seemingly attractive solution is to postulate that dark matter
interacts with a light force carrier $\phi$ with fine structure
constant $\alpha_X \equiv
\lambda^2/(4\pi)$~\cite{Cirelli:2008pk,ArkaniHamed:2008qn}.  For
$m_{\phi} = 0$, the annihilation is enhanced by the Sommerfeld
factor~\cite{Sommerfeld:1931}
\begin{equation}
S = \frac{\pi \, \alpha_X / \vrel}{1 - e^{-\pi \alpha_X / \vrel}} \ .
\label{Sbar}
\end{equation}
For massive $\phi$, the enhancement is typically cutoff at a value
$\propto \alpha m_X/
m_{\phi}$~\cite{Hisano:2002fk,Cirelli:2007xd,Cirelli:2008pk,%
ArkaniHamed:2008qn}, but for fine-tuned choices of $\alpha_X$, $m_X$
and $m_\phi$, there are also resonance regions where the enhancement
may exceed this cutoff, as we discuss below.  The velocity of dark
matter particles is $\sim 1/3$ at freeze out and $\sim 10^{-3}$ now.
Sommerfeld enhancement therefore provides an elegant mechanism for
boosting annihilations now.  Constraints from dark matter annihilation
in protohalos with $\vrel \sim 10^{-8}$ exclude
$m_\phi=0$~\cite{Kamionkowski:2008gj}.  However, taking $m_X \sim
\tev$ and $m_\phi \sim \mev - \gev$, and assuming $\langle \sigmaan
\vrel \rangle \approx \sigmath$, one may still generate $S \sim 10^3$
to explain the positron excesses, while the cutoff allows one to
satisfy the protohalo constraint.

Of course, for a viable solution, dark matter must not only annihilate
with the correct rate, it must also be produced with the right density
and form structure in accord with observations.  Here we find that the
desired thermal relic density cannot be achieved in
Sommerfeld-enhanced models designed to explain the positron excesses.
In addition, we show that the new force carrier $\phi$ induces dark
matter self-interactions that may contradict current observations.  As
is well-known, if $\phi$ were massless, the resulting long range
Coulomb force would lead to large energy transfers that make halos
spherical, and observations of triaxial halos constrain this
possibility~\cite{Ackerman:2008gi,Feng:2009mn}.  For $m_\phi \sim
100~\mev$, the force's range is only $\sim 10~\fm$, but, as we show
below, the implied cross section is still large enough to play a role
in galactic dynamics.

\ssection{Thermal Relic Density}
If $XX$ annihilation is enhanced by $\phi$ exchange, there is an
``irreducible'' annihilation process $XX \to \phi \phi$ through 
$t$-channel $X$.  For $m_\phi \ll m_X$, the thermally-averaged
annihilation cross section is the typical WIMP cross section 
\begin{equation}
\langle \sigmaan \vrel \rangle \approx \pi\alpha_X^2 / m_X^2 \ ,
\label{sigmaann}
\end{equation}
with ${\cal O}(1)$ corrections depending on the details of the initial
and final states.  Requiring that $\langle \sigmaan \vrel \rangle$ be
small enough that $X$ can be all of the dark matter implies
\begin{equation}
\alpha_X \leq \sqrt{\sigmath / \pi} \ m_X  \ .
\label{alphaconstraint}
\end{equation}
This bound is conservative.  In fact, the Sommerfeld effect enhances
the annihilation cross section even at freeze
out~\cite{Cirelli:2008pk,Dent:2009bv}, and the bound may be
significantly strengthened in the presence of other annihilation
channels.

\ssection{Self-Interactions} 
Self-interactions allow dark matter particles to transfer energy.  The
average rate for dark matter particles to change velocities by ${\cal
O}(1)$ factors is
\begin{equation}
\Gamma_k=\int d^3v_1 d^3v_2 f(v_1) f(v_2) \left(n_X \vrel \sigma_T \right)
\left(\vrel^2 / v_0^2\right) ,    
\end{equation}
where $f(v) = e^{-v^2/v^2_0} / (v_0\sqrt\pi)^3$ is the dark matter's
assumed (Maxwellian) velocity distribution, $n_X$ is its number
density, $\vrel=|\vec{v}_1-\vec{v}_2|$, and $\sigma_T = \int
d\Omega_\ast (d\sigma/d\Omega_\ast)(1- \cos \theta_\ast)$ is the
energy transfer cross section, where $\theta_\ast$ is the scattering
angle in the center-of-mass frame.

Dark matter particles coupled to a massive force carrier $\phi$
scatter through the Yukawa potential $V(r)=- \alpha_X e^{-m_\phi r} /
r$.  In the Born approximation, keeping only the dominant $t$-channel
contribution present in all interactions, the transfer cross section
is
\begin{equation}
\sigma_T = \frac{2 \pi}{m_\phi^2} \beta^2 
\left[ \ln  \left( 1 + R^2 \right) - \frac{R^2}{1 + R^2} \right] ,
\label{sigmatransfer}
\end{equation}
where $\beta \equiv 2 \alpha_X m_\phi/(m_X \vrel^2)$ is the ratio of
the potential energy at $r\sim m_\phi^{-1}$ to the kinetic energy of
the particle, and $R \equiv m_X \vrel / m_\phi$ is the ratio of the
interaction range to the dark matter particle's de Broglie
wavelength. For typical values of interest here, $\vrel \sim 10^{-3}$
and $m_X/m_\phi \agt 10^3$, and so $R \agt 1$.  For $R \gg 1$,
$\sigma_T \approx \frac{8 \pi \alpha_X^2}{\vrel^4 m_X^2} \left(\ln R^2
- 1 \right)$.  As in the Coulomb case, this is greatly enhanced for
small $\vrel$, but here the finite interaction length of the Yukawa
potential cuts off the logarithmic divergence.

\Eqref{sigmatransfer} receives significant corrections in the strong
interaction regime, where $\beta \gg 1$.  Our focus in this work will
be on the $R \gg 1$ region of parameter space. In this region, quantum
effects are subdominant and so classical studies~\cite{Khrapak:2003}
of particles moving in Yukawa potentials are applicable.  Although the
authors of these studies were interested in slow and highly charged
particles moving in plasmas with screened Coulomb potentials, they
approximated these potentials by Yukawa potentials, and so their
results are exactly applicable in the current context.  The numerical
results of these studies are accurately reproduced
by~\cite{Khrapak:2003}
\begin{eqnarray}
\sigma_T &\simeq& \frac{4 \pi}{m_\phi^2} \beta^2 \ln \left( 1 +
\beta^{-1} \right)\ , \quad \beta < 0.1 \ , \nonumber \\
\sigma_T &\simeq& \frac{8 \pi}{m_\phi^2} \frac{\beta^2}{1 + 1.5
  \beta^{1.65}} \ , \quad 0.1 < \beta < 1000 \ .
\label{sigmaTclassical}
\end{eqnarray}
We use these analytical fits to obtain the results below.

\ssection{Halo Shapes}
Self-interactions that are strong enough to create ${\cal O}(1)$
changes in the energies of dark matter particles will isotropize the
velocity dispersion and create spherical halos. These expectations are
borne out by simulations of self-interacting dark matter in the hard
sphere limit~\cite{Spergel:1999mh,Dave:2000ar,Yoshida:2000bx}.  The
shapes of dark matter halos of elliptical galaxies and clusters are
decidedly elliptical, which constrains
self-interactions~\cite{MiraldaEscude:2000qt}.  The ellipticity of
galactic halos provides the strongest constraints on these
models~\cite{Feng:2009mn}. To implement these constraints, we consider
the well-studied, nearby (about 25 Mpc away) elliptical galaxy NGC
720.  In Ref.~\cite{Buote:2002wd}, X-ray isophotes were used to
extract the ellipticity of the underlying matter
distribution. Comparing it to the ellipticity induced by the stellar
mass profile, the dark matter halo of NGC 720 was found to be
elliptical at about 5 kpc and larger radii.

To compute $\Gamma_k$, we use the measured total mass profile and the
decomposition into stars plus dark matter for NGC
720~\cite{Humphrey:2006rv} and obtain the radial velocity dispersion
$\overline{v^2_r}(r)=v^2_0(r)/2$ and the dark matter density. For the
radius we pick $5~\kpc$. Our constraints would be stronger if we could
use the higher densities inside this radius, but the constraints on
the ellipticity weaken for radii below 5 kpc~\cite{Buote:2002wd}. For
the dark matter density, we choose the average value within 5 kpc,
which is roughly $4~\gev/\cm^3$. To compute the dispersion, we assume
isotropy and that the total (stellar plus dark matter) mass profile
scales approximately linearly with radius. For an NFW profile with
best fit scale radius~\cite{Humphrey:2006rv},
$\overline{v^2_r}(r)\simeq (240~\km/\s)^2$. Varying within the quoted
error range for the scale radius~\cite{Humphrey:2006rv} only changes
this dispersion by about 10\%.

\begin{figure}[tb]
\begin{center}
\includegraphics*[width=0.92\columnwidth]{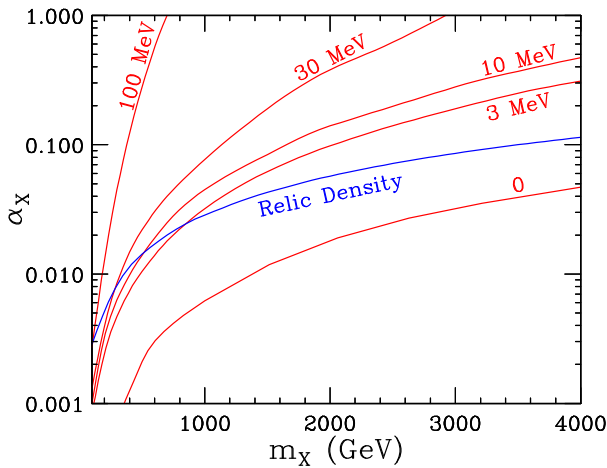}
\end{center}
\vspace*{-.25in}
\caption{Regions above the contours are excluded by the relic density
constraint and by halo ellipticity observations for the $m_\phi$
indicated. The classical approximation used to obtain the halo bounds
becomes inaccurate for $m_\phi \agt 100~\mev$.
\label{fig:alphamxmphi}}
\end{figure}

\ssection{Results}
To derive constraints on the particle physics parameters from the
observed halo shapes, we require
\begin{equation}
\Gamma_k^{-1} > 10^{10}~\text{years} \ ,
\label{constraint}
\end{equation}
\ie, that the average time for self-interactions to create ${\cal
O}(1)$ changes in dark matter particle velocities is greater than the
galaxy's lifetime.  Imposing \eqsref{alphaconstraint}{constraint} from
the relic density and the observation of ellipticity in the dark
matter halo of NGC 720 yields the constraints shown in
\figref{alphamxmphi}.  The relic density constraint is independent of
$m_\phi$ and the extremely stringent halo shape constraint for $m_\phi
= 0$~\cite{Feng:2009mn} remains significant for $m_\phi \alt 30~\mev$.
The crucial point is that when the interaction range is larger than
the de Broglie wavelength, although the Coulomb logarithm enhancement
is lost, the enhancement from low $\vrel$ remains.  Note that our
assumption of a locally Gaussian velocity distribution is supported by
recent simulation of Milky Way-sized dark matter
halos~\cite{Vogelsberger:2008qb}.  $\Gamma_k$ does not change by more
than a factor of about 2 when we allow the distribution to become
anisotropic or introduce a velocity cut-off at the escape speed.  At
the same time, we have checked that our halo shape bounds are
consistent with the predictions from simulations with hard sphere
scattering~\cite{Dave:2000ar}.

\begin{figure}[tb]
\begin{center}
\includegraphics*[width=0.92\columnwidth]{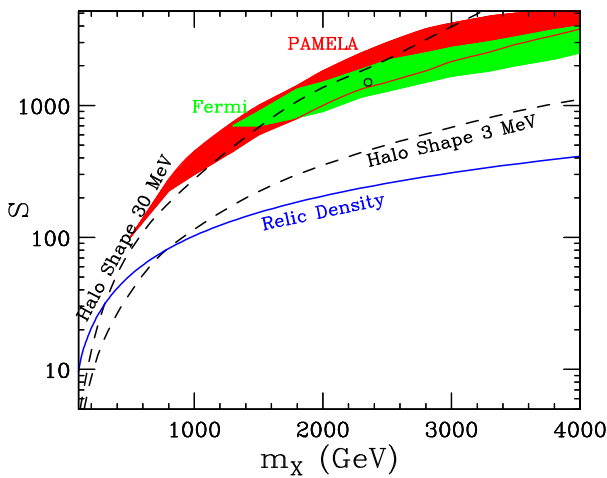}
\end{center}
\vspace*{-.25in}
\caption{Upper bounds on Sommerfeld enhancement factor $S$ from relic
density (solid), along with PAMELA- and Fermi-favored regions and the
best fit point $(m_X, S) = (2.35~\tev, 1500)$~\cite{Bergstrom:2009fa},
all for $m_{\phi} = 250~\mev$.  Halo shape bounds are also shown for
the values of $m_\phi$ indicated (dashed).
\label{fig:Sbounds}}
\end{figure}

In \figref{Sbounds} we present the regions of the $(m_X, S)$ plane
required to explain PAMELA and Fermi as determined in
Ref.~\cite{Bergstrom:2009fa}.  These are for $m_\phi = 250~\mev$,
which is large enough to allow contributions to positrons through
$\phi \to \mu^+ \mu^-$, but small enough to forbid contributions to
anti-protons, where no excess is seen~\cite{ArkaniHamed:2008qn}.
Upper bounds from relic density and halo shapes are also given.  We
see that the large Sommerfeld enhancements required to explain the
positron excesses are significantly excluded by the relic density
constraint for all $m_X$.  For $m_\phi \alt 30~\mev$, the halo shape
constraints also exclude the required Sommerfeld enhancements.

\ssection{Discussion}
The results of \figref{Sbounds} are not surprising.  For the relic
density, the WIMP miracle implies that for $m_X \sim 250~\gev$, the
correct relic density is obtained for $\alpha \sim 10^{-2}$.  Given
$\vrel \sim 10^{-3}$, this implies an upper bound of $S \sim 10$, and
this bound scales as $m_X$.  Of course, $X$ need not be all the dark
matter, but in this case, the Sommerfeld-enhanced flux scales as $n^2
\langle \sigmaan v \rangle S \sim \alpha_X^{-1}$, and so the signal is
maximal for $S \sim 1$.

In deriving our results, we have ignored the cutoff of the Sommerfeld
enhancement factor for massive $\phi$.  Including this cutoff will
reduce the maximal possible $S$ for low $m_X$, strengthening the
disagreement between the allowed values of $S$ and the experimentally
favored regions. To reduce the disagreement, one might consider
resonant Sommerfeld enhancement. As with resonances from additional
postulated particles~\cite{Feldman:2008xs}, these resonances require
fine-tuning and are bounded by astrophysical
observations~\cite{Profumo:2009uf}.  In addition, resonance
enhancement occurs at $m_\phi / m_X \simeq 6 \alpha_X / (\pi^2 n^2)$,
$n = 1, 2, \ldots$~\cite{Cassel:2009wt} and is significant only for
low $n$.  For $m_\phi \sim \gev$ and the relevant range of $\alpha_X
\agt 0.01$, this implies $m_X \alt 500~\gev$; the resonances are,
therefore, ineffective in reaching the favored regions given in
\figref{Sbounds}.  Most importantly, as noted above, our bounds are
conservative in that they do not include the Sommerfeld effect on
freeze out~\cite{Dent:2009bv}.  This effect suppresses the largest
possible $S$, especially at resonances.  Self-consistently including
the effects of resonances on annihilation in both the early Universe
and now, we find that the maximal possible enhancement factor is $S
\sim 100$, even allowing for resonances~\cite{inprep}.

As an alternative approach to evade the relic density constraints, one
may consider other production mechanisms or modify early Universe
cosmology, but this sacrifices the WIMP miracle and also removes the
motivation for considering Sommerfeld enhancement in the first place.
Alternatively, one might appeal to boosts of $\sim 10$ from cold and
dense dark matter substructure in the local neighborhood. Such large
values at a distance of only 10 kpc from the Milky Way center are,
however, not motivated by simulations with collisionless dark
matter~\cite{Vogelsberger:2008qb}. The presence of the stellar disk
would further reduce these expectations. The self-scatterings among
the particles in the substructure would also serve to reduce the inner
densities~\cite{Dave:2000ar} and hence the expected boost. In
addition, for $m_\phi \alt 30~\mev$, interactions with the dark matter
particles of the Milky Way could evaporate substructure because
$\vrel$ is much larger than the internal velocity dispersion of the
substructure~\cite{Gnedin:2000ea}.

The halo shape bounds are obtained from inferred dark matter halo
ellipticity, which depends on merger histories and the
environment. For example, a major merger at a redshift of $z=0.5$ for
NGC 720 would effectively halve the age that $\Gamma_k^{-1}$ should be
compared to and weaken the bound on $m_\phi$ by roughly a factor of
$\sqrt{2}$. However, the lack of large scale disturbances in the gas
argues against such a recent major merger. These bounds may be made
more robust by deeper data sets of NGC 720, which will further
constrain point source contamination and rotation or large scale
disturbances in the gas, as well as by measuring ellipticities and
mass profiles in other galaxies and clusters~\cite{Buote:1997ab}.

A second prediction of strongly self-interacting dark matter is the
formation of constant density cores, if gravo-thermal collapse does
not occur.  The time scale for the formation of these cores is of
order $\Gamma_k^{-1}$, suggesting that NGC 720 should have ${\cal
O}(\kpc)$ sized core.  Future tests for the presence of cores in
galaxy and cluster halos may provide comparable or stronger limits.
Self-interactions should also dramatically alter the dark matter halos
of smaller galaxies, such as the dwarf galaxies in the Local
Group. The central dark matter densities measured in these dwarf
satellites of the Milky Way are ${\cal
O}(\gev/\cm^3)$~\cite{Strigari:2008ib} and fit neatly within the
standard CDM predictions. For the parameter space disfavored by NGC
720 observations, and using simulation results~\cite{Dave:2000ar}, we
estimate that core sizes would be of order the luminous extent of the
dwarfs or larger.  The tidal force of the Milky Way would
significantly reduce the central densities of the dwarfs with such
large cores and likely make it impossible to explain the large
observed dark matter densities in all the
dwarfs~\cite{Penarrubia:2010jk}.  In parameter regions with more
moderate Sommerfeld enhancements, these cores would be smaller and
consistent with current data~\cite{Gentile:2004tb}.

\ssection{Conclusions}
Cosmic positron data have motivated dark matter candidates with
Sommerfeld-enhanced annihilations.  The required enhancement is large,
requiring large couplings to light force carriers.  Annihilation to
these force carriers provides an upper limit on the thermal relic
abundance of these dark matter candidates. With or without resonances,
this constraint excludes the existence of enhancements that can
explain the positron excesses. These models also predict
self-interactions that may make galactic dark matter halos
spherical. The ellipticity of the halo of NGC 720 also excludes the
required Sommerfeld enhancements for $m_{\phi} \alt 30~\mev$.
Interestingly, viable models with moderate Sommerfeld enhancements,
although unable to explain the positron data, may predict constant
density spherical cores in small galactic halos and other departures
from the standard cold dark matter paradigm that are consistent with
current data.

\ssection{Acknowledgments} 
We thank Matthew Buckley, David Buote, Patrick Fox, Phil Humphreys,
Masahiro Ibe, Alessandro Strumia, and Huitzu Tu for helpful
conversations. The work of JLF and HY was supported in part by NSF
grants PHY--0653656 and PHY--0709742. The work of MK was supported in
part by NSF grant PHY--0855462 and NASA grant NNX09AD09G.

\ssection{Note added}
As the first version of this work was being completed, we learned of
related work in progress. This work~\cite{Buckley:2009in} agrees with
\eqref{sigmaTclassical} in the classical regime.



\end{document}